\documentclass[seceq]{ptptex}
\usepackage{amsmath}
\usepackage[dvips]{graphicx}
\usepackage{amssymb}

\def\R{{\mbox{\boldmath $R$}}}
\def\Z{{\mbox{\boldmath $Z$}}}

\def\mcl{\rm}

\title{
On the Topology of Black Lenses
}

\author{
Daisuke {\sc Ida}
}

\inst{
Department of Physics, Gakushuin University, Tokyo 171-8588, Japan
}

\recdate{
April 23, 2009
}

\abst{
The topological structure of the black holes
in 5-dimensional space-times
with a horizon diffeomorphic with the lens space
has been discussed.
It has been shown that such a black hole can emerge from 
the crease set,
which is composed of the plumbings of several 2-spheres,
of the event horizon.
It has also been shown that such configurations are realized
in the Kastor-Traschen solution of the Einstein-Maxwell
system.
}

\begin{document}

\maketitle

\section{Introduction}
The subject of the present article is the topological structure of the black holes
in 5-dimensional space-times with horizons diffeomorphic with lens spaces.
The first conclusive result on the topology of the black holes is due to 
Hawking~\cite{Hawking:1971vc},
who shows under the dominant energy condition
that  apparent horizons in  4-dimensional space-times
must be diffeomorphic with the 2-sphere, or possibly with the 2-torus.
The possibility of the torus horizon is later excluded by 
topological censorship theorem.~\cite{Friedman:1993ty,Jacobson:1994hs,Browdy:1995qu}

The Hawking's theorem can be generalized in 5 or greater space-time dimensions as follows;
under the dominant energy condition,
the apparent horizon admits a metric with positive scalar curvature, or it is Ricci flat.
The Ricci flat case, which corresponds to the torus horizon case in 4-dimensions,
 has not been excluded without stronger assumptions.
This possibility, however, seems implausible, so that we will not consider the Ricci flat
horizon in what follows.
In 5 space-time dimensions, it follows that the allowed horizon topology is
$S^3$, $S^2\times S^1$, the lens spaces $L(p,q)$, and the connected sum of 
finite numbers of their copies.~\cite{Cai:2001su,Helfgott:2005jn,Galloway:2005mf}

Among these, Emparan and Reall~\cite{Emparan:2001wn} 
show that the black hole with $S^2\times S^1$ horizon, which is usually called
the {\em black ring}, is realized as 
the solution to the vacuum Einstein equation
in the 5-dimensional, stationary and
asymptotically flat space-time.
The black hole with the lens space horizon in general
will be called the {\em black lens}.
It has never been known whether a black lens is realized as a stationary and
asymptotically flat solution 
to the Einstein equation or not.~\cite{Evslin:2008gx,Chen:2008fa}. \
However, it is easily constructed with looser asymptotic conditions.
Since a lens spaces is the quotient space $S^3/\Gamma$ 
of the 3-sphere
with a finite subgroup $\Gamma$
of $SO(4)$ acting freely on $S^3$,
a black lens can be obtained as the quotient space of any spherically symmetric,
in the sense that $SO(4)$ acts isometrically on $S^3$,
black hole space-time, say, of the 5-dimensional Schwarzschild space-time.

For the existence of the black lens, the underlying space should have nontrivial
topology, for the black lens cannot be embedded into $\R^4$ in general.
Hence, the topology of the event horizon itself is also topologically nontrivial.
To understand the topological structure of the black lens space-times, 
the quotient space model stated above is not so helpful.
Let us consider the formation of a black lens via the gravitational collapse
within the quotient space model.
A simple model is obtained from the 5-dimensional Oppenheimer-Snyder model
via the identification of points in the $S^3$ part 
of the space-time geometry.
Then, we will not have a regular Cauchy surface in the quotient space-time. 
In other words, we will rather have a conical singularity at the symmetric center
before, or at the early stage of, the gravitational collapse.
Roughly speaking, the topological information has been stuffed into the conical singularity.
The main purpose of the present article is to clarify the topological structure of
event horizon in 5-dimensional space-times with the black lenses.

This paper is organized as follows.
In Sec.~\ref{sec:preliminaries},
the mathematical preliminaries are reviewed,
where
basic facts on the topological structure of event horizons
are summarized, and
the lens space and the handle attachment are briefly introduced.
In Sec.~\ref{sec:lp1},
the special lens space of type $L(p,1)$ is described as the boundary of
the 2-disk bundle over the 2-sphere.
In Sec.~\ref{sec:gls},
the general lens space of type $L(p,q)$ and the 4-manifold, which has $L(p,q)$
as the boundary,  are described.
In Sec.~\ref{sec:lsbh},
a topological model of the event horizon of the black lens is proposed.
In Sec.~\ref{sec:kt},
we show that the 5-dimensional Kastor-Traschen space-time models contain a realization
of our model.
The section~\ref{sec:conc}
gives conclusion.

\section{Preliminaries}\label{sec:preliminaries}

\subsection{Event horizons and crease sets}
\begin{figure}[htbp]
\centerline{\includegraphics[width=.6\linewidth]{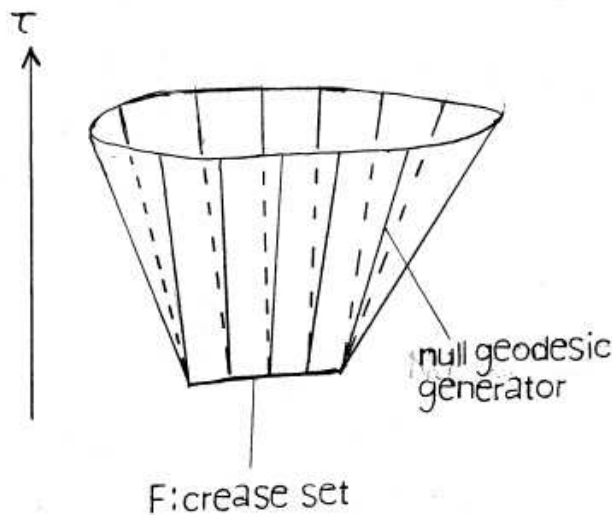}}
\caption{The crease set $F$ is the subset of the event horizon consisting of all the
past end points of the null geodesic generators the event horizon.}
\label{fig:1_crease_set}
\end{figure}

The event horizon in $(n+1)$-dimensional space-time $M$ is a $n$-dimensional submanifold $H$.
It is generated by null geodesics in $M$ without future end points.\cite{HE}
The set consisting of all the past end points of 
the null geodesic generators of $H$ will be called
the {\em crease set}~\cite{Siino:1997ix} and it will be denoted by $F$ 
(See  Fig.~\ref{fig:1_crease_set}).
In other words, the crease set $F$ is the spot in the space-time 
from which the event horizon emanates.
Clearly, the crease set $F$ is a subset of $H$.
Furthermore, $F$ is an achronal set, for $H$ is a null hypersurface.
The event horizon $H$ is in general not smoothly embedded in $M$, but 
it rather has a wedge-like structure at the crease set $F$.

Suppose that there is a global time function $\tau: M\to \R$ in $M$,
such that $\langle X,d\tau\rangle >0$ holds for all future pointing nonzero 
non-space-like vector $X$.
Since each null geodesic generator of $H$ intersects $\tau={\rm const.}$ hypersurface at most once,
a point in $H$ is locally parametrized by $(\tau,x^i)$ $(i=1,\cdots,n-1)$,
where $(x^i)$ parametrize null geodesic generators of $H$.
In what follows, we mainly consider situations where $H$ is within the portion
$\tau\ge \tau_0$ for some real constant $\tau_0$.
In other words, we are interested in black holes formed by gravitational collapses.

The portion of the event horizon $H\cap \{\tau\le \tau_1\}$ $(\tau_1={\rm const.})$ will be
denoted by $W(\tau_1)$.
The set $W(\tau_1)$ will be a compact set.
The boundary $\dot{W}(\tau_1)$ of $W(\tau_1)$
is called the black hole boundary, or simply the horizon, at the time $\tau=\tau_1$.
The topology of the horizon $\dot{W}(\tau)$ will in general change with the time $\tau$.
The topology change of the horizon occurs only when the time slice
$\tau={\rm const.}$ intersects the crease set $F$.

Each null geodesic generator $(x^i)={\rm const.}$ has a past end point in $F$
at the time $\tau=\tau_F(x^i)$. Then, the $\epsilon$-neighborhood $N_\epsilon (F)$
of $F$ in $H$ is defined for a positive number $\epsilon$ 
to be the union 
of all the null geodesic segments $\{(\tau,x^i)\in H;\tau_F(x^i)\le\tau< \tau_F(x^i)+\epsilon\}$
of the null geodesics generators $(x^i)={\rm const.}$ of $H$.
Since the null geodesic generators of $H$ does not have any caustics apart from $N_\epsilon(F)$,
the boundary $\dot{N}_{\epsilon}(F)$ of $N_\epsilon(F)$ will be homeomorphic with
the horizon $\dot{W}(\tau)$ at sufficiently late times.
In this sense, we can say that the topological information of the event horizon $H$ is 
encoded in its crease set $F$.

Here we give a more definitive characterization of the crease set,
which seems to have never appeared in the literature.
To state this, let me recall several terminologies in elementary topology.
Let $X$ be a topological space and let $A$ be a subset of $X$.
A homotopy $f:X\times [0,1]\to X$ is called a {\em deformation retraction},
if 
$f(X,0)={\rm id}_X$, $f(X,1)\subset A$, and $f(A,1)={\rm id}_A$
hold, and the subset $A$ is called a {\em deformation retract} of $X$.
In this terminology, the following statement holds;
{\em the crease set $F$ is a deformation retract of the event horizon $H$.}
This can be seen by introducing the homotopy $f:H\times[0,1]\to H$ by
\begin{eqnarray*}
  f[(\tau,x^i),t]&=&\left\{
\begin{array}{cc}
(\tau,x^i)& (\tau\le T(t))\\
(\max\{T(t),\tau_F(x^i)\},x^i)& (\tau>T(t))
\end{array}
\right.\\
T(t)&=&\tan\left[{\pi(1-2t)\over 2}\right]
\end{eqnarray*}
It is easily seen that the homotopy $f$ gives a deformation retraction of $H$
onto $F$.

A deformation retract $A$ of $X$ has the same homotopy type
with $X$.
Therefore, all the homotopy groups of the crease set $F$ are isomorphic with those of 
the event horizon $H$.
In particular, $F$ is arcwise connected if $H$ is, therefore the black hole horizon 
at late times is, arcwise connected.

\subsection{Lens spaces}\label{subsec:Lens_Space}
Let us for a moment recall how to construct  lens spaces.
The lens space is a closed 3-manifold obtained by identification of boundaries
of a pair of solid toruses.
A solid torus $V$ is a product space $S^1\times D^2$. 
Let us introduce the coordinate system on $V$ as follows.
Let $(\rho,\phi)$, $\rho\in[0,1]$, $\phi\in [0,2\pi)$ be the polar coordinates
on $D^2$ and let $\theta\in [0,2\pi)$
be the coordinate on $S^1$, where the coordinates $\phi$ and $\theta$
are defined modulo $2\pi$. The coordinates $(\rho,\phi,\theta)$ parametrize
points on $V$. 
The simple closed curves $m$ and $l$
given by $(\rho,\phi,\theta)=(1,t,0)$, $t\in[0,2\pi]$,
and by $(\rho,\phi,\theta)=(1,0,t)$, $t\in[0,2\pi]$,
in the boundary torus $\partial V\simeq T^2$ of $V$ are called meridian and longitude
of $V$,
respectively.
Any simple 
closed curve in $\partial V$ is homotopic to a curve $pl+qm$, where 
$p$ and $q$ are coprime integers or $(p,q)=(0,0)$.

Let $V'\simeq S^1\times D^2$ be a copy of $V$ and let
$(\rho',\phi',\theta')$ be the corresponding coordinate system
on $V'$, and let $m'$ and $l'$ be the meridian and the longitude of
$V'$, respectively.
The boundary of $V$ ($V'$) is parametrized by the coordinates 
$(\phi,\theta)$ $[(\phi',\theta')]$.
Identifying points in $\partial V$ and points in $\partial V'$,
a closed 3-manifold is obtained.
The identification is defined by a diffeomorphism  
$f: \partial V\to \partial V'$ between two toruses.
The 3-manifold obtained by this identification will be denoted by 
$V\cup_f V'$.
Let the orientation of $V$ be given such that the coordinate system
$(\rho,\phi,\theta)$ is positively ordered,
and let the orientation of 
$V\cup_fV'$ be induced from $V$.
The diffeomorphism type of $V\cup_f V'$ is determined, but not uniquely, 
by the homotopy class of $f$
within self-diffeomorphisms of $T^2$.
The homotopy class of $f$ is represented by the map of the form
\begin{eqnarray*}
  \left(\begin{array}{c}\phi\\\theta\end{array}\right)
=
A
  \left(\begin{array}{c}\phi'\\\theta'\end{array}\right)
=\left(\begin{array}{cc}q&r\\p&s\end{array}\right)
  \left(\begin{array}{c}\phi'\\\theta'\end{array}\right)
\end{eqnarray*}
where $A$ belongs to $GL(2,\Z)$.
In particular, $p$ and $q$ are coprime for $qs-pr=\pm 1$ holds.
A pair $(p,q)$ of coprime integers determines the homotopy class $[f(m')]$ of
the image of the meridian $m'$ on $\partial V$. 
In fact, for the parametrization above of $A$, $[f(m')]$ is written as
\begin{eqnarray*}
  [f(m')]=p[l]+q[m]\label{eq:L(p,q)}
\end{eqnarray*}
where $[\gamma]$ denotes the homotopy class of the closed curve $\gamma$ 
in $\partial V$.
For fixed coprime pair $(p,q)$, there are infinitely many pairs of integers
$(r,s)$ such that $qs-pr=\pm 1$ holds. 
The diffeomorphism type
of $V\cup_f V'$ is determined, though still not uniquely, by $(p,q)$, and
it does not depend on the choice of $(r,s)$.
This is seen as follows. Let $D$ be the embedded 2-disk in $V'$ bounded by $m'$ and let $N(D)$
be the tubular neighborhood of $D$ in $V'$. Let $\overline f:
\partial V'\cap \partial N(D)\to \partial V$ be the restriction of $f$
to $\partial V'\cap \partial N(D)$.
It is clear that the 3-manifold $V\cup_{\overline f} N(D)$ is determined by a pair $(p,q)$
irrespective of the choice of $(r,s)$. 
The boundary of $V\cup_{\overline f} N(D)$ is the 2-sphere,
and the closed manifold $V \cup_f V'$ is obtained 
from $V\cup_{\overline f} N(D)$ and the 3-disk $V'-N(D)$
by identifying their boundary 2-spheres. However, it is known that closed 3-manifold
$X_1$ and $X_2$ are diffeomorphic if 
$X_1-{\rm int} D^3$ and $X_2-{\rm int}D^3$ are diffeomorphic,
where $X_i-{\rm int}D^3$ denotes the complement of the embedded open 
3-disk in $X_i$, for $i=1,2$.
It follows that  $V\cup_f V'$ is uniquely determined by $V\cup_{\overline f} N(D)$
and hence only by a pair $(p,q)$.

The oriented closed 3-manifold $V\cup_f V'$ characterized by Eq.~(\ref{eq:L(p,q)})
is denoted by $L(p,q)$ and it is called the {\em lens space of type $(p,q)$}.
Note that we can always take $p\ge 0$ for $L(p,q)\simeq L(-p,-q)$ follows
from the definition.
Note also that $L(p,q)\simeq L(p,-q)$ by orientation reversing diffeomorphism.
It is known that two lens spaces $L(p,q)$ and $L(p',q')$ $(p,p'\ge 0)$
are diffeomorphic,
if and only if (i) $p=p'$ and (ii) $q=\pm q'$ $({\rm mod}~p)$ or
$qq'=\pm 1$ $({\rm mod}~p)$ hold.

\subsection{2-handle attachments}
In the author's previous work~\cite{Ida:2007ti} on the event horizon,
the elementary process of the topology change of the black holes is 
understood as the handle attachment, where building blocks of the event horizon
are played by $m$-handles ($0\le m\le n-1$) for $(n+1)$-dimensional space-times.
In our model of the black lens, only the 2-handle attachment is relevant, 
so that we explain the 2-handle attachment to some extent.
We consider the 5-dimensional black hole space-times.
Suppose that the event horizon changes its topology at the time $\tau=\tau_1$.
Then, the topology change is described by a handle attachment.
Let this be described by the 2-handle attachment.

\begin{figure}[htbp]
\centerline{\includegraphics[width=.7\linewidth]{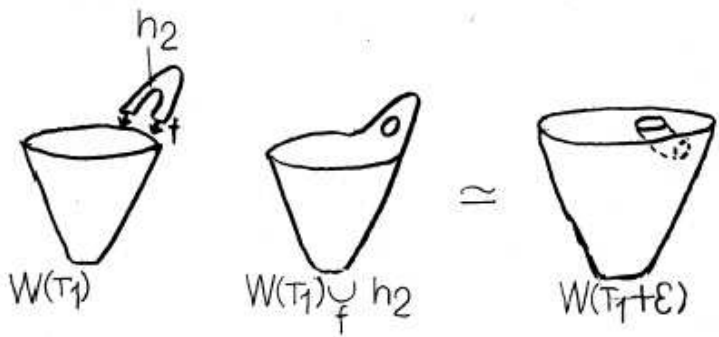}}
\caption{The topology change of the event horizon can be regarded as the
handle attachment.
The attachment of the 2-handle $h_2$ to the 4-dimensional 
event horizon $W(\tau_1)$ at the time $\tau=\tau_1$
is schematically depicted. In this figure, 2 dimensions are omitted, where $h_2$ becomes
$D^1\times D^1$ instead of $D^2\times D^2$.}
\label{fig:2_2handle_attachment}
\end{figure}

The 2-handle is the 4-disk regarded as the product space $D'\times D''$,
where both $D'$ and $D''$ are topological 2-disks, and it is denoted by $h_2$.
The part $\tau\le \tau_1$ of the event horizon $W(\tau_1)$ changes its topology
by attaching the 2-handle $h_2$ along its boundary, which is called
the {\em 2-handle attachment}. The 2-handle attachment is specified by the 
embedding
$f:\partial D'\times D''\to \partial W(\tau_1)$, of the solid torus
into the boundary of $W(\tau_1)$,
which is called the {\em gluing map}
The new 4-space is obtained from
the disjoint union $W(\tau_1)\sqcup h_2$ 
via the identification $x\sim f(x)$ for all $x\in \partial D'\times D''$,
and it is denoted by $W(\tau_1)\cup_f h_2$
(See Fig.~\ref{fig:2_2handle_attachment}).

The part of the event horizon $W(\tau_1+\epsilon)$ 
for a small positive number $\epsilon$ is by definition homeomorphic with $W(\tau_1)\cup_f h_2$.

\section{$L(p,1)=S^3/\Z_p$ horizons}\label{sec:lp1}
The most simple lens space black hole evolves from a crease set homeomorphic with $S^2$.
The subject of this section is the black hole of which spatial section is diffeomorphic
with special lens space $L(p,1)$ at late times.
Since the lens space cannot be embedded in $\R^4$, the space-time which has lens space horizon
must be topologically nontrivial.
Then, the $\epsilon$-neighborhood of the crease set in the event horizon
may not be the direct product space
$S^2\times D^2$, but in general is $D^2$ bundle over $S^2$.

The $D^2$ bundle over $S^2$ is constructed in the following way.
Let $U_1$ and $U_2$ be a pair of the unit 2-disks, and consider the
product bundles $B_1=U_1\times F_1$ and $B_2=U_2\times F_2$,
where $F_1$ and $F_2$ are the unit 2-disks.
The coordinate system on $B_i$  is given by 
$(r_i,\theta_i,\rho_i,\phi_i)$,
 ($r_i$, $\rho_i\le 1$),
such that
$(r_i,\theta_i)$ is the polar coordinate system on $U_i$ and 
$(\rho_i,\phi_i)$ the polar
coordinate system 
on $F_i$, for $i=1,2$.
The $D^2$ bundle over $S^2$ is obtained from $B_1$ and $B_2$ via
the identification as follows; a point $(r_1,\theta_1,\rho_1,\phi_1)$ on $B_1$
and a point $(r_2,\theta_2,\rho_2,\phi_2)$ on $B_2$ are identified if
\begin{eqnarray}
r_1&=&r_2=1\\\nonumber
\theta_1&=&\theta_2\\\nonumber
\rho_1&=&\rho_2\\\nonumber
\phi_1&=&\phi_2+p\theta_2
\label{eq:attaching_map(p,1)}
\end{eqnarray}
hold, where $p\in \Z$.
It is known that any $D^2$ bundle over $S^2$ is equivalent with one 
those constructed above, hence it is characterized by an integer $p$.
\begin{figure}[htbp]
\centerline{\includegraphics[width=.5\linewidth,height=.3\textheight]{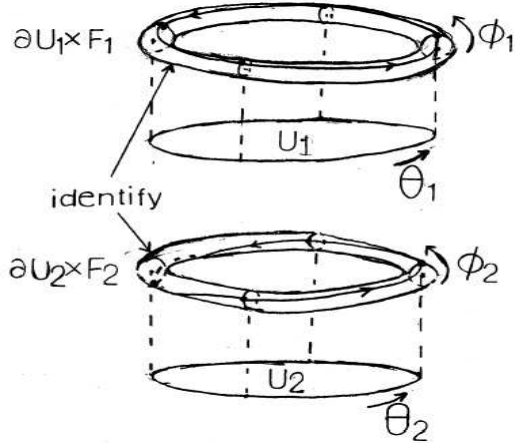}}
\caption{The construction of the $D^2$ bundle over $S^2$.
The parts $\partial U_i\times F_i$ $(i=1,2)$ of 
the product spaces $U_i\times F_i$  are depicted.
The $D^2$ bundle over $S^2$ is obtained 
from this pair of the product spaces
via the identification between 
the solid toruses $\partial U_1\times F_1$
and  $\partial U_2\times F_2$.
This figure illustrates $p=1$ case, where the
identification is made such that 
the closed curve drawn on the torus $\partial U_1\times \partial F_1$
is identified with  the closed curve drawn on the torus $\partial U_2\times \partial F_2$
}
\label{fig:3_disk_bundle}
\end{figure}
This integer $p$ is called the Euler number of the associated $\R^2$ bundle.
Let $M_p$ denote the $D^2$ bundle over $S^2$ so constructed.

The boundary $\partial M_p$ of the 4-manifold $M_p$ is diffeomorphic
with the lens space $L(p,1)$.
This can be seen as follows.
The boundary of the product space $U_i\times F_i$
is composed of
a pair of solid toruses
$V_i=U_i\times \partial F_i$ and 
$\widetilde V_i=\partial U_i\times F_i$,
and
\begin{eqnarray*}
\partial B_i=V_i\cup \widetilde V_i\\
V_i\cap \widetilde V_i=\partial V_i=\partial\widetilde V_i
\end{eqnarray*}
hold,
for $i=1,2$.
Noting that $M_p$ is 
the disjoint union $B_1\sqcup B_2$
with the identification 
$f: V_2 \to \widetilde V_1$
defined by Eqs.~(\ref{eq:attaching_map(p,1)}),
it is seen that
\begin{eqnarray*}
  \partial M_p=V_1\cup \widetilde V_2\\
V_1\cap \widetilde V_2=\partial V_1=\partial \widetilde V_2
\end{eqnarray*}
hold in $M_p$. In other words, $\partial M_p$ is the closed
3-manifold obtained by gluing boundaries of a pair of solid toruses
$V_1$ and $\widetilde V_2$.
Hence, $\partial M_p$ is a lens space.
To see that $\partial M_p\simeq L(p,1)$, it is sufficient to note that
$f$ maps the meridian $\widetilde m_2$ of $\widetilde V_2$ 
onto the closed curve homotopic to
$pl_1+m_1$ in $\partial V_1$, 
where  $l_1$ and $m_1$ are the longitude and
the meridian of $V_1$, respectively.
In fact, $\widetilde m_2$ is written as
$(r_2,\theta_2,\rho_2,\phi_2)=(1,t,1,0)$ $(t\in[0,2\pi])$,
and its image under $f$ is
$(r_1,\theta_1,\rho_1,\phi_1)=(1,t,1,pt)$, $(t\in[0,2\pi])$,
which is homotopic to $pl_1+m_1$ in $\partial V_1$.

The 4-manifold $M_p$ can be considered as the topological model of the event horizon which
becomes lens space black hole diffeomorphic with $L(p,1)\simeq S^3/\Z_p$ at late times.
In fact, $M_p$ is the tubular neighborhood of 
$U_1\cup U_2\simeq S^2$
and the boundary $\partial M_p$ is the $S^1$ bundle over $S^2$, which is
diffeomorphic with $L(p,1)$. 
Then,
the crease set of the event horizon 
can be regarded as
corresponding 
to the base space $S^2$ of $M_p$.

The 4-manifold $M_0$ is considered as the model of the event horizon which evolves into
the black ring. Since $M_0$ is the product bundle $S^2\times D^2$, its boundary is
$S^2\times S^1$. This example shows that there are at least two topologically distinct 
processes in the formation of a black ring. 
The first example, which  is perhaps most intuitive one,
is the gravitational collapse 
of the massive object with the shape of  a circle $S^1\times D^3$
with finite thickness in the 5-dimensional Minkowski background.
When the gravitational collapse occurs first in the inward direction of $D^3$, the
black ring will be formed, and the event horizon will emanate from the $S^1$ crease set.
The another example, which might be less intuitive, 
is given by the present model $M_0$. Consider the gravitational collapse
of the massive object with the shape of a  2-sphere
$S^2\times D^2$ with finite thickness in the 5-dimensional Minkowski
background space-time. When the $D^2$ part collapses first, the black ring will again be formed.
In this time, the crease set will be a 2-sphere.
Roughly speaking, the former example corresponds to the formation of a thin black ring
in the sense that the radius of the $S^1$ is larger than that of $S^2$
and the latter that of a thick one vise versa.

On the other hand,
the 4-manifold $M_1$ can be considered as the model of the event horizon which has the 
$L(1,1)\simeq S^3$ 
spatial section at late times. The $S^1$ bundle over $S^2$, $\partial M_1=L(1,1)$,
is diffeomorphic with $S^3$, which is known as the Hopf fibration of $S^3$.
This example shows that the event horizon evolving into an $S^3$ black hole need not be
homeomorphic with 4-disk. In fact, $M_1$ cannot be a 4-disk, since it has an incompressible
2-sphere $U_1\times \{0\}\cup U_2\times \{0\}$ as its submanifold.
This situation can be contrasted with that in the 4-dimensional general relativity
as follows.
In 4-dimensional space-times, the event horizon $H$, which evolves into an $S^2$ black hole,
must be the topological 3-disk and the crease set must be contractible to a point in $H$.

In general, the boundary of $M_p$ is the lens space $L(p,1)$. Hence $M_p$ can be regarded
as the topological model of the event horizon which evolves into a black hole
with $L(p,1)$ horizon at late times.
The construction above of $M_p$ is exactly the 2-handle attachment to $B_1$, when
$B_2$ is viewed as the 2-handle $h_2$. Thus, the formation of the $L(p,1)$ horizon
can be regarded as the 2-handle attachment to the $D^4$ event horizon.

\section{A construction of general lens space}\label{sec:gls}

In the previous section,
the formation of the $L(p,1)$ black hole 
is described as a 2-handle attachment to a 4-disk.
We show in the next section that 
a topological model of formation of
$L(p,q)$ horizons for general $(p,q)$
is obtained by several 2-handle attachments to a 4-disk.
For this purpose, let us for a moment make purely 3-dimensional argument on the boundary of 
the event horizon.

\subsection{Dehn surgery}
\begin{figure}[htbp]
\centerline{\includegraphics[width=.5\linewidth]{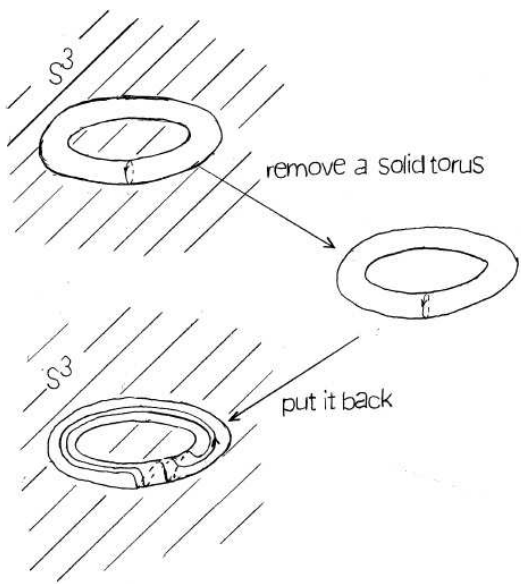}}
\caption{The Dehn surgery in $S^3$ is depicted. The solid torus is put back
such that the meridian is mapped onto the closed curve indicated in the figure.}
\label{fig:4_Dehn_surgery}
\end{figure}

Consider  the effect on the boundary $\partial M$ of the 4-manifold $M$
  caused by the 2-handle attachment.
It removes a solid torus $V$ from $\partial M$, 
and puts another solid torus $V'$ back,
by gluing boundary torus $\partial(\partial M-{\rm int}V)$ and $\partial V'$ together.
In general, the operation,
which  removes a solid torus from a 3-manifold and  glues another solid torus back
along the boundary toruses,
is called the {\em Dehn surgery} (See Fig.~\ref{fig:4_Dehn_surgery}).
Noting that the complement of an open solid torus in the 3-sphere
is again a solid torus, it is understood that 
the general lens space is obtained by performing the Dehn surgery once in the 3-sphere.

Let $V$ be the solid torus embedded in the 3-manifold $X$ without boundary
and let $V'$ be another solid torus.
Let $l$ ($l'$) and $m$ ($m'$) be the longitude and the meridian of $V$ ($V'$), respectively.
The Dehn surgery of $X$ is determined by the
diffeomorphism $f:\partial V'\to \partial V$ of the torus.
The resultant 3-manifold is the disjoint union $(X-{\rm int}V)\sqcup V'$ 
with the identification defined by $f$, and it will be denoted by 
$X'=(X-{\rm int}V)\cup_f V'$. Let the orientation of $X'$ be that induced from  $X$.
By argument in Sec.~(\ref{subsec:Lens_Space}),
the diffeomorphism type of $X'$ is determined by the homotopy class $[f(m')]$ 
in $\partial V$
of the image of the meridian of $V'$ under $f$.
It is written as
\begin{eqnarray*}
  [f(m')]=p[m]+q[l]
\end{eqnarray*}
in terms of coprime integers $p$ and $q$. 
Then, the rational number 
$r=p/q$ ($r=\infty$, when $q=0$) is called the {\em surgery coefficient}.
Note that the change of the boundary associated with the 2-handle attachment to the 4-manifold
corresponds to the Dehn surgery with the integral surgery coefficient.
Note also that, when $q=0$, 
$p$ must be unity for $f(m')$ to be a simple closed curve in $\partial V$,
and the resultant 3-manifold $X'$ is diffeomorphic with $X$.
In other words, the Dehn surgery with $r=\infty$ has no effect.

\begin{figure}[htbp]
\centerline{\includegraphics[width=.3\linewidth]{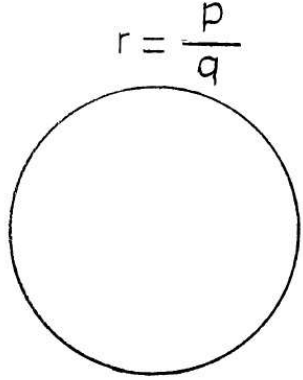}}
\caption{The Dehn surgery is depicted as a circle with a coefficient.}
\label{fig:5_circle}
\end{figure}

It is convenient to introduce a pictorial representation of the Dehn surgery.
The Dehn surgery is depicted as the  circle in the closed 3-manifold $X$
with the coefficient.
In the same notation above,
let the orientation of the solid torus on $V$ be 
that induced from $X$.
Regarding $V$ as $D^2\times S^1$, introduce the coordinate system $(\rho,\phi,\theta)$,
$\rho\in[0,\epsilon]$, $\phi\in[0,2\pi)$, $\theta\in[0,2\pi)$
on $V$, such that $(\rho,\phi)$ is the polar coordinate system on $D^2$ and $\theta$ 
parametrizes
$S^1$ and that the coordinates $(\rho,\phi,\theta)$ are positively ordered.
In terms of the coordinates, the longitude $l$ of $V$ is written as 
$(\rho,\phi,\theta)=(\epsilon, 0,t)$, $t\in[0,2\pi]$ and the meridian $m$
of $V$, $(\rho,\phi,\theta)=(\epsilon,t,0)$, $t\in[0,2\pi]$.
Then, the Dehn surgery is depicted as the  circle $(\rho,\phi,\theta)=(0,0,t)$,
$t\in[0,2\pi]$, which is the central circle of $V$ with the surgery coefficient $r=p/q$ 
written near the circle. 
(Fig.~\ref{fig:5_circle})

If we take $X=S^3$, a single circle labeled by $r=p/q$ represents the lens space $L(p,q)$.
This is seen from the expression $X'=(S^3-{\rm int}V)\cup_f V'$,
noting that  $S^3-{\rm int}V$ is the solid torus.
In general, all closed orientable 3-manifold can be 
obtained by Dehn surgeries associated with some 
link in the 3-sphere 
consisting of circles with coefficients. 
The diagram of such link in the 3-sphere is called the {\em Kirby diagram}~\cite{Kirby}
of the corresponding closed orientable 3-manifold.

\subsection{The Hopf link in 3-sphere}
\begin{figure}[htbp]
\label{fig:6_Hopf_link}
\centerline{\includegraphics[width=.5\linewidth]{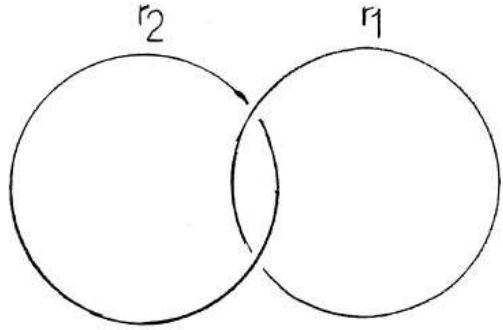}}
\caption{The Hopf link in 3-sphere.}
\end{figure}

\begin{figure}[htbp]
\centerline{\includegraphics[width=.5\linewidth]{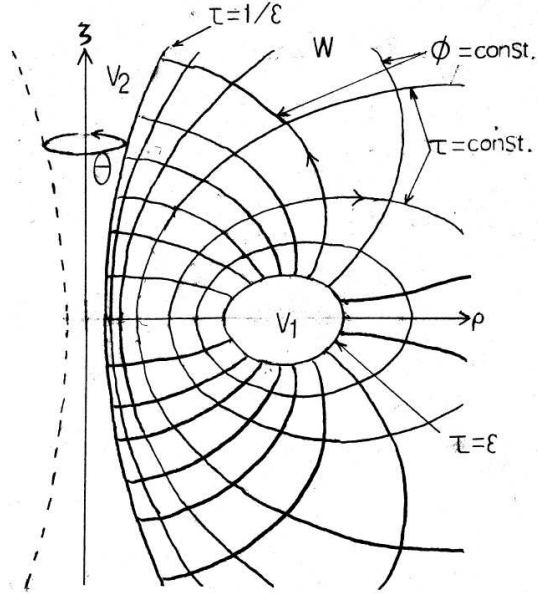}}
\caption{The torus coordinate system $(\tau,\phi,\theta)$ is illustrated in $\R^3$.
The regions $V_1$ and $V_2$ indicated in the figure
are solid toruses in $S^3=\R^3\cup \{\infty\}$.}
\label{fig:7_bipolar_coordinates}
\end{figure}

\begin{figure}[htbp]
\centerline{\includegraphics[width=.6\linewidth]{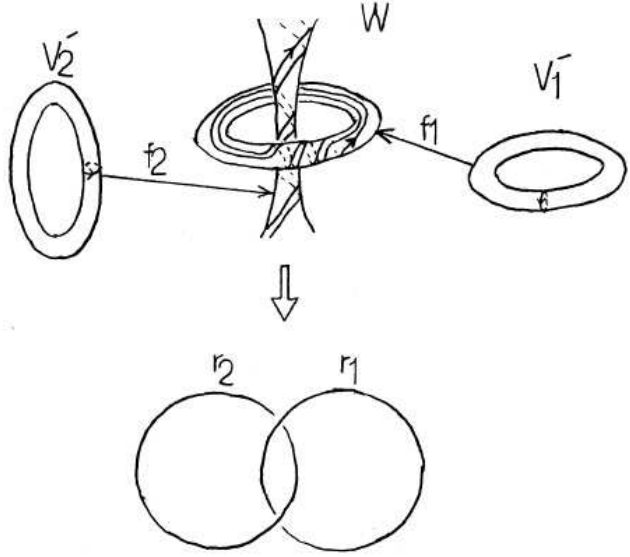}}
\caption{The 3-manifold $W'=(W\cup_{f_1}V_1')\cup_{f_2}V_2'$ is represented as
the Hopf link in $S^3$.}
\label{fig:8_Hopf_surgery}
\end{figure}

In the next three subsections, we will see that all lens spaces can be 
obtained by several Dehn surgeries in the 3-sphere only with integer surgery coefficients.
The number of the integer Dehn surgeries required depend on the type $(p,q)$ of lens space.
In a simple case when the required number of the integer 
Dehn surgeries is two, it is represented by
a pair of circles linked together exactly once, which is usually called
the {\em Hopf link}, in the 3-sphere
(Fig.~\ref{fig:6_Hopf_link}).
To study the Hopf link in the 3-sphere, 
let us introduce the coordinate system in the 3-sphere.
Since 
$S^3-\{\mbox{a point}\}\simeq \R^3$ ,
points in $S^3-\{\mbox{a point}\}$
can be parametrized by the Euclidean coordinates $(x,y,z)$.
The preferred coordinate system here is the {\em torus coordinates}
$(\tau,\phi,\theta)$, which we call.
The torus coordinate system is defined
in terms of the cylindrical coordinates $(\rho,\vartheta,\zeta)$
\begin{eqnarray*}
  x=\rho \cos \vartheta,\quad
  y=\rho\sin\vartheta,\quad
z= \zeta
\end{eqnarray*}
by
\begin{eqnarray*}
\tau&=&
{(\rho-1)^2+\zeta^2\over\rho}\\
  \phi&=&\left\{\begin{array}{cc}
\dfrac{\pi}{ 2}\left[1+\dfrac{
\rho^2+\zeta^2-1}
{\sqrt{(\rho^2+\zeta^2-1)^2+4\zeta^2}}\right]&
(\zeta\ge 0)\\
-\dfrac{\pi}{ 2}\left[1+ \dfrac{\rho^2+\zeta^2-1}{\sqrt{(\rho^2+\zeta^2-1)^2+4\zeta^2}}
\right]&
(\zeta<0)\end{array}\right.\\
\theta&=&\vartheta\\
\end{eqnarray*}
where the coordinates $\phi$ and $\theta$  are identified modulo $2\pi$.
Each surface $\tau={\rm const}.$ is a torus parametrized by 
$(\phi,\theta)$ (See Fig.~\ref{fig:7_bipolar_coordinates}).
Consider the region $W:\{\epsilon<\tau<1/\epsilon\}$, where
$\epsilon$ is a positive number strictly less than unity.
The torus coordinate system $(\tau,\phi,\theta)$ is regular in $W$.
The closure of the complement $S^3-W$ 
of $W$ in $S^3$ consists of two solid toruses $V_1$ and $V_2$
with the central
circles $C_1:\{(\rho,\zeta)=(1,0)\}$ and $C_2:\{\rho=0\}$, respectively, in terms of the cylindrical
coordinates.
The meridian $m_1$ and the longitude $l_1$ of the solid torus $V_1$ are given by
\begin{eqnarray*}
&m_1&: (\tau,\phi,\theta)=(\epsilon,0,t) \quad(t\in [0,2\pi])\\
&l_1&: (\tau,\phi,\theta)=(\epsilon,t,0) \quad(t\in [0,2\pi])
\end{eqnarray*}
respectively,
while the meridian $m_2$ and the longitude $l_2$ of the solid torus
$V_2$ are given by
\begin{eqnarray*}
&m_2&: (\tau,\phi,\theta)=(1/\epsilon,t,0) \quad(t\in [0,2\pi])\\
&l_2&: (\tau,\phi,\theta)=(1/\epsilon,0,t) \quad(t\in [0,2\pi])
\end{eqnarray*}
respectively.
Let $V_1'$ and $V_2'$ be solid toruses,
and let $m_1'$ and $m_2'$ be the meridian of $V_1'$ and $V_2'$, respectively.
Let $f_i:\partial V_i'\to \partial V_i$ $(i=1,2)$
be diffeomorphism of torus.
Then, the Dehn surgery is determined by
$W'=(W\cup_{f_1} V_1')\cup_{f_2} V_2'$, which is the disjoint union
$W\sqcup V_1'\sqcup  V_2'$ with identification given by $f_1$ and $f_2$.
The diffeomorphism type of $W'$ is determined by the homotopy classes
$[f_1(m_1')]$ and $[ f_2( m_2')]$ in $\partial W$.
They can be written as
\begin{eqnarray*}
  [f_1(m_1')]&=&p_1[m_1]+q_1[l_1]\\
{}[f_2(m_2')]&=&p_2[m_2]+q_2[l_2]
\end{eqnarray*}
The Dehn surgery determined by $f_1$ and $f_2$ is represented as two circles linked together,
one with coefficient $r_1=p_1/q_1$ and the other with $ r_2=p_2/ q_2$.
(See Fig.~\ref{fig:8_Hopf_surgery})
\subsection{$\mcl S$-transformation}
Here we show that the same 3-manifold is obtained, if the pair $(r_1, r_2)$ of the surgery
coefficients are replaced by the pair $(-1/ r_2,-1/r_1)$.
To see this, consider the following embedding of $W$ into the 3-sphere.
Let $\overline {S}$ be another 3-sphere,
and let $(\overline{\vphantom{\phi}\tau},\overline\phi,\overline\theta)$
denote the torus coordinate system on $\overline S$.
Similarly, the copy of $V_1$, $V_2$, $m_1$, $l_1$, $m_2$, $l_2$ in $\overline S$
are denoted
as $\overline V_1$, $\overline V_2$, 
$\overline m_1$, $\overline l_1$, 
$\overline m_2$, $\overline l_2$,
respectively.
Define the embedding $\psi: W\to \overline S$ by
\begin{eqnarray*}
  \overline \tau&=&1/\tau\\
\overline\phi&=&\phi\\
\overline\theta&=&-\theta
\end{eqnarray*}
This transformation exchanges the roles of the meridian and the longitude.
In other words,
\begin{eqnarray*}
\psi(m_1)&=&\overline{l}_2\\
 \psi(l_1) &=&-\overline{m}_2\\
 \psi(m_2)&=&-\overline l_1\\
 \psi (l_2)&=&  \overline m_1
\end{eqnarray*}
hold, where we use the same notation $\psi$ for the map between oriented curves 
induced by $\psi$.
Therefore, the homotopy classes $[f_1(m_1')]$ and $[ f_2(m_2')]$
in $\partial W$ are mapped to the homotopy classes
\begin{eqnarray*}
{}  [\widetilde{f_2}( m_1')]&=&p_1[\psi(m_1)]+q_1[\psi(l_1)]
=p_1[\overline{l}_2]-q_1[\overline{m}_2]\\
{}[\widetilde{f_1}( m_2')]&=&p_2[\psi(m_2)]+q_2[\psi(l_2)]
=- p_2[\overline l_1]+ q_2[\overline m_1]
\end{eqnarray*}
in $\partial W$, respectively,
where $\widetilde f_1=\psi\circ f_2$ and $\widetilde{ f_2}=\psi\circ f_1$
are defined.
Clearly, $\overline W'=(\psi(W)\cup_{\overline{f}_1}V_2')\cup_{\overline{f}_2} V_1'$
is diffeomorphic with $W'$.
Hence, the pair of surgery coefficients $(r_1,r_2)$ can be replaced by
the pair $(-1/r_2,-1/r_1)$ without affecting the diffeomorphism type of the corresponding
3-manifold.
Let us call this operation for a Hopf link  ${\mcl S}$.
\begin{eqnarray}
(r_1,r_2)\xrightarrow{\mcl S}\left(-\dfrac{1}{r_2},-\dfrac{1}{ r_1}\right)  
\end{eqnarray}

\subsection{${\mcl T}^n$-transformation}
Next we show that the linked circles with coefficients $(r_1,r_2)$
can be replaced also by $(r_1+n,(n+1/r_2)^{-1})$ for $n\in \Z$.
Consider the embedding $\sigma:W\to \overline S$
given by
\begin{eqnarray*}
\overline \tau&=&\tau\\
\overline \phi&=&\phi+n \theta\\
\overline\theta&=&\theta
\end{eqnarray*}
where $n$ is an integer.
By this transformation, the images of $m_i$ and $l_i$ $(i=1,2)$ are
\begin{eqnarray*}
 \sigma (m_1)
&=&\overline m_1\\
\sigma (l_1)
&=&n \overline m_1+\overline l_1\\
\sigma (m_2)
&=&\overline m_2+n \overline l_2\\
\sigma (l_2)
&=&\overline l_2
\end{eqnarray*}
respectively.
Therefore, the homotopy classes $[f_1(m_1')]$ and $[f_2(m_2')]$ in $\partial W$
are mapped to the homotopy classes
\begin{eqnarray*}
{}  [\widehat f_1(m_1')]&=&p_1[\sigma (m_1)]+q_1[\sigma(l_1)]
=(p_1+n q_1)[\overline m_1]+q_1[\overline l_1]\\
{}[\widehat f_2(m_2')]&=& p_2[\sigma (m_2)]+q_2[\sigma (l_2)]
=p_2[\overline m_2]+(n p_2+q_2)[\overline l_2]
\end{eqnarray*}
in $\partial \overline W$, respectively,
where $\widehat f_1=\sigma\circ f_1$ and $\widehat f_2=\sigma\circ f_2$ are defined.
It follows that the pair $(r_1,r_2)$ of surgery coefficients can be replaced by
the pair $(r_1+n,(n+1/r_2)^{-1})$ without affecting the diffeomorphism type
of the corresponding 3-manifold.
Let us call this transformation of surgery coefficients ${\mcl T}^n$.
\begin{eqnarray}
(r_1,r_2)\xrightarrow{{\mcl T}^n}\left(r_1+n,\dfrac{1}{n+\dfrac{1}{ r_2}}\right)  
\end{eqnarray}

\subsection{General lens space}
\begin{figure}[htbp]
\centerline{\includegraphics[width=.5\linewidth]{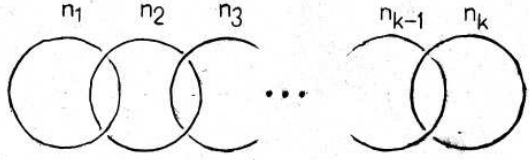}}
\caption{The lens space of type $(p,q)$ is represented by the Kirby diagram
of linearly aligned linked circles with integral coefficients.}
\label{fig:9_general_lens}
\end{figure}

The general lens space is represented by the
finite number of linearly aligned linked circles each with an integral coefficient as in 
Fig.~\ref{fig:9_general_lens}.
To see this, let us verify a procedure to eliminate a single circle from the links.
First, consider a pair of linked circles one with integral coefficient $n$ and the other
with rational coefficient $r$.
According to the setting in the present section,
let the circles labeled by $n$ and $r$
be located at $\tau=0$ and $\tau=\infty$, respectively.
This corresponds to the pair of surgery coefficients $(n,r)$.
This pair is equivalent with the pair $(-1/r,-1/n)$ via the
${\mcl S}$-transformation, and further it is equivalent with
the pair $(n-1/r,\infty)$ via the successive ${\mcl T}^n$-transformation.
\begin{eqnarray*}
  (n,r)\xrightarrow{\mcl S} \left(-\dfrac{1}{r},-\dfrac{1}{n}\right)
\xrightarrow{\mcl T^{n}} 
\left(n-\dfrac{1}{r},\infty\right)
\end{eqnarray*}
This is equivalent with the single circle with rational coefficient
$n-1/r$,
for the circle with a coefficient $\infty$ can be eliminated.

\begin{figure}[htbp]
\centerline{\includegraphics[width=.5\linewidth]{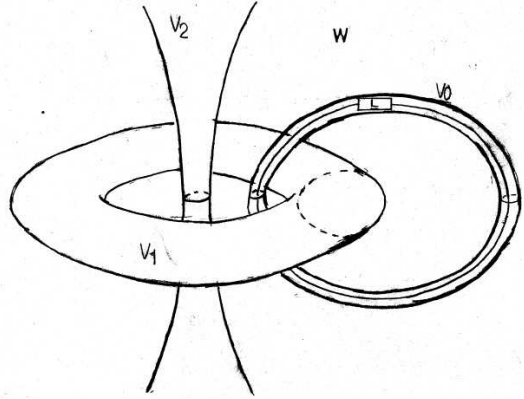}}
\caption{The location of the solid torus $V_0$ in $W$ is depicted.
The structure of links within $V_0$,  denoted by $L$ in the figure,
 does not change
under the ${\mcl S}$ and ${\mcl T}^n$-transformations.}
\label{fig:10_V0}
\end{figure}

Next, let us consider the situation, where other components consisting of 
 labeled circles, which are denoted by $L$ as a whole,
are present besides the pair $(n,r)$ above.
Assume that any labeled circle in $L$ is not linked with the circle labeled by $r$. 
In other words, 
the existence of
an embedded 2-disk in the 3-sphere,
which is bounded by the circle labeled by $r$ and does not intersect $L$,
is assumed.
Let the circles labeled by $n$ and $r$ be located at $\tau=0$ and
$\tau=\infty$, respectively, as above.
Then, we can assume that $L$ is entirely located in the interior of the thin solid torus
$V_0$ determined by $(1-\tau)^2+\theta^2\le\delta$, where $\delta$ is a small
positive number such that $V_0$ is within the interior of $W$
(See Fig.~\ref{fig:10_V0}).
Let us see the effects of ${\mcl S}$ and ${\mcl T}^n$-transformations on $L$.
Note that all points on the central circle $C_0$ of $V_0$
given by $(\tau,\phi,\theta)=(1,t,0)$, $t\in[0,2\pi]$ are 
fixed points under ${\mcl S}$ and ${\mcl T}^n$. 
The effect of ${\mcl S}$-transformation on $V_0$ 
is essentially a composition of 
reflections in $\tau$ and $\theta$ directions with respect to $C_0$, 
which is just  simultaneous $180^\circ$ rotations of the meridian disks
given by $\phi={\rm const.}$ of $V_0$.
Therefore, the structure of $L$ remains unchanged under ${\mcl S}$-transformation.
The effect of ${\mcl T}^n$-transformation on $V_0$ is  distortion of
the annuluses given by $\tau={\rm const.}$ in terms of the 
element of $SL(2,\R)$ in the $(\phi,\theta)$-plane. 
Hence, the structure of $L$ remains unchanged also under ${\mcl T}^n$.

\begin{figure}[htbp]
\centerline{\includegraphics[width=.5\linewidth,height=.3\textheight]{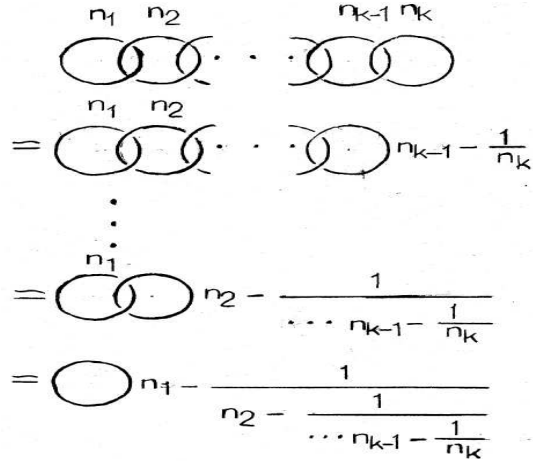}}
\caption{The procedure  reducing linked circles into a single circle is illustrated.
Linked circles in every line represent the same 3-manifold.}
\label{fig:11_Kirby_move}
\end{figure}

We are now in position to confirm that linearly aligned
linked circles, as in Fig.~\ref{fig:9_general_lens}, each labeled by integer
are equivalent with a single labeled circle with a rational coefficient.
Let us represent these linked circles by the ordered surgery coefficients
$(n_1,n_2,\cdots,n_k)$.
Identify the pair of circles labeled by $(n_{k-1},n_k)$ with the pair $(n,r)$ above,
and regard the remaining circles $(n_1,\cdots,n_{k-2})$ as $L$.
Then, the ${\mcl S}$ followed by ${\mcl T}^{n_{k-1}}$ results in
linked circles $(n_1,\cdots,n_{k-2},n_{k-1}-1/n_k)$ consisting of fewer circles by one.
By recursive procedures, we end at the single circle with the coefficient
\begin{eqnarray*}
  r=n_1-\dfrac{1}{n_2-\dfrac{1}{n_3-\dfrac{1}{\cdots n_{k-1}-\dfrac{1}{n_k}}}}
\end{eqnarray*}
This procedure is called the {\em rational blow down} in the literature.
Conversely, for given a circle labeled by a rational number $r$, 
there is an equivalent Kirby diagram consisting of
a finite number of linearly aligned linked circles with integral coefficients.
Hence, any lens space can be constructed by several Dehn surgeries with integral coefficients
(See Fig.~\ref{fig:11_Kirby_move}).

\section{The model of black lenses}\label{sec:lsbh}
Let us consider the 4-manifold which has the lens space as its boundary
as a model describing lens space black holes.
As explained in Secs.~\ref{sec:lp1} and~\ref{sec:gls},
the Kirby diagram of a single circle with integer coefficient $p$ represents 
the lens space $L(p,1)$
 and it is realized as the boundary of 
the 2-disk bundle $M_p$ over the 2-sphere.
Correspondingly, $M_p$ is constructed by the 2-handle attachment to the 4-disk.
The construction is as following.
The Kirby diagram is regarded as that in the boundary 3-sphere in the 4-disk.
The attaching circle for the 2-handle attachment is the circle, denoted by $C$, of
the Kirby diagram.
Let $V$ be the solid torus, which is the tubular neighborhood of $C$ in $\partial D^4$
and let $V'$ be the solid torus $D^2\times \partial D^2$ in the boundary of $h_2$,
when $h_2$ is written as $D^2\times D^2$.
Then, the attaching map $f$ is the diffeomorphism of $V'$ onto $V$,
and it is determined
by the integer coefficient $p$ labeling the circle,
in such a way that the image of the longitude $l'$ of $V'$ under $f$
is homotopic to the curve $l+pm$ in $\partial V$, 
\begin{eqnarray*}
  [f(l')]=[l]+p[m]
\end{eqnarray*}
where $l$ and $m$ denote
the longitude and meridian of $V$, respectively.
This 
is precisely equivalent with the construction of $M_p$ explained in Sec.~\ref{sec:lp1}

Next, let us consider the 4-manifold corresponding to the Kirby diagram of
the Hopf link consisting of the circles $C$ and $C'$ with integer 
coefficients $p$ and $p'$, respectively.
It is constructed by 2-handle attachments twice to the 4-disk.
This 4-disk can be considered as the product space $D^2\times D^2$.
Let us introduce the coordinate system $(r,\theta,\rho,\phi)$ in $D^2\times D^2$, where
$(r,\theta)$ $r\in[0,1]$, $\theta\in[0,2\pi)$ 
[{\it resp.} $(\rho,\phi)$, $\rho\in[0,1]$, $\phi\in[0,2\pi)$] is the polar coordinate system in $D^2$ of
the first ({\it resp.} second) factor of $D^2\times D^2$.
The attaching circles $C$ and $C'$ can be regarded as being located at
$C:\{r=1,\rho=0\}$ and $C':\{r=0,\rho=1\}$, respectively, where two circles
$C$ and $C'$ are linked together exactly once in the boundary 3-sphere of $D^2\times D^2$.
To see that $C$ and $C'$ together form a Hopf link in $\partial (D^2\times D^2)$,
note that $C'$ is homotopic to the circle $C'':\{r=1,\phi=0,\rho=1\}$ 
in $\partial (D^2\times D^2)-C$. 
Clearly, $C''$ bounds the 2-disk $\{r=1,\phi=0\}$ in $\partial(D^2\times D^2)$,
and intersects $C$ exactly once.
This implies that $C$ and $C''$, and hence $C$ and $C'$,
 form a Hopf link in $\partial (D^2\times D^2)$.

\begin{figure}[htbp]
\centerline{\includegraphics[width=.5\linewidth]{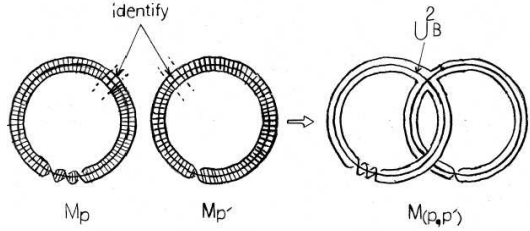}}
\caption{The construction of the plumbing manifold is schematically  illustrated,
where  $D^2$ bundles over $S^2$ are drawn as  $D^1$ bundles over $S^1$.}
\label{fig:12_plumbing}
\end{figure}

First, the 2-handle is attached along $C$. 
Then, we obtain the 2-disk bundle $M_p$ over
the 2-sphere as explained in this section above.
The coordinates $(r,\theta)$ parametrize (the half of) the base space
of $M_p$
and  $(\rho,\phi)$ parametrize its fibre.
Next, the 2-handle is attached to $M_p$ along $C'$.
The resultant 4-manifold will be denoted by $M_{(p,p')}$.
However, we can also exchange the order of these 2-handle attachments.
If we performed the first 2-handle attachment along $C'$, we would obtain
the 2-disk bundle $M_{p'}$ first. 
Then, the coordinates $(r,\theta)$ would parametrize
the fibre of $M_{p'}$ and $(\rho,\phi)$ would parametrize (the half of) its base space.
From this observation, the 4-manifold $M_{(p,p')}$ turns out to be
the disjoint union $M_p\sqcup M_{p'}$ with identification given by the 
gluing map
\begin{eqnarray*}
g:U\times F\to U'\times F';\quad
(x,y)\mapsto (x',y')=(y,x)  
\end{eqnarray*}
where $U$ ($U'$) is the 2-disk embedded in the base space of $M_p$ ($M_{p'}$)
and $F$ ($F'$) is its 2-disk fibre.
In other words, the diffeomorphism $g$ identifies the fibre $F$ of $U\times F$ with
the base space $U'$ of $U'\times F'$, and the base space $U$ of $U\times F$ with the
fibre $F'$ of $U'\times F'$.
This type of operation to patch a pair of 2-disk bundles over the 2-sphere together,
is called the {\em plumbing}, and the 4-manifold obtained by the plumbing
is called the plumbing manifold
(See Fig.~\ref{fig:12_plumbing}).
The plumbing manifold $M_{(p,p')}$ is no longer a fibre bundle.
The boundary of $M_{(p,p')}$ is the lens space $L(pp'-1,p')$,
which is implied by the transformations 
\begin{eqnarray*}
  (p,p')\xrightarrow{\mcl S}(-1/p',-1/p)
\xrightarrow{{\mcl T}^p} (p-1/p',\infty)
\end{eqnarray*}
of the surgery coefficients performed in the boundary.
Let $B$ and $B'$ be the 2-spheres as the base spaces of the fibre bundles
$M_p$ and $M_{p'}$, respectively.
Consider the subset $B\cup_{\overline g} B'$ of $M_{(p,p')}$,
which is the disjoint union of $B$ and $B'$ with identification
given by the restriction $\overline g$ of $g$ to the point $(0,0)\in U\times F$.
In other words, this set is a pair of 2-spheres with a point in $B$ and
a point in $B'$ identified. 
Roughly speaking, $B\cup_{\overline g}B'$ is thought of as the frame of $M_{(p,p')}$.
The 4-manifold $M_{(p,p')}$ can be regarded as the model of an event horizon
which evolves into a black hole with the horizon diffeomorphic with $L(pp'-1,p')$.
The crease set of the event horizon corresponds to $B\cup_{\overline g}B'$,
since $M_{(p,p')}$ is the $\epsilon$-neighborhood of $B\cup_{\overline g}B'$.

Now, we are in position to give a model of an event horizon, which evolves 
into a black hole with the horizon diffeomorphic with $L(p,q)$ for
general coprime integers $p$ and $q$.
First, consider the continued fraction of $p/q$.
Then, we will have
\begin{eqnarray*}
  {p\over q}=
n_1-\dfrac{1}{n_2-\dfrac{1}{\cdots n_{k-1}-\dfrac{1}{n_k}}}
\end{eqnarray*}
for some $k\in Z$.
This implies that $L(p,q)$ is represented by the Kirby diagram of
linearly linked circles
with integer coefficients $(n_1,n_2,\cdots,n_k)$.
Let $g_i$ be the gluing map giving the plumbing between the fibre bundles
$M_{n_i}$ and $M_{n_{i+1}}$. 
The 4-manifold 
\begin{eqnarray*}
M_{(n_1,n_2,\cdots,n_k)}=M_{n_1}\cup_{g_1}M_{n_2}\cup_{g_2}\cdots \cup_{g_{k-1}}M_{n_k}  
\end{eqnarray*}
is constructed from $k$ fibre bundles via the successive plumbings,
where we assume that the image of $g_i$ and the preimage of $g_{i+1}$ have no intersection
for all $i=1,\cdots k-1$.
The frame of $M_{(n_1,\cdots ,n_k)}$ will be denoted by
\begin{eqnarray*}
\cup^k B=B_1\cup_{\overline g_1}\cdots \cup_{\overline g_k}B_k
\end{eqnarray*}
where $B_i$ is the base space of $M_{n_i}$ $(i=1,\cdots,k)$.
The structure of $\cup^k B$ itself depends only on the number $k$ of the length of the
continued fraction, for it is the disjoint union of 2-spheres $B_1,\cdots,B_k$
with a point $x_i\in B_i$ and a point $y_{i+1}\in B_{i+1}$ identified, respectively
for $i=1,\cdots,k-1$, where $x_{i}\ne y_i$.
Clearly, the boundary of $M_{(n_1,\cdots,n_k)}$ is the lens space $L(p,q)$, and
$M_{(n_1,\cdots,n_k)}$ is the $\epsilon$-neighborhood of $\cup^k B$.
Hence, the plumbing manifold
$M_{(n_1,\cdots,n_k)}$ gives a model of an event horizon which evolves
into a lens space black hole with the horizon diffeomorphic with $L(p,q)$,
and the subset $\cup^k B\subset M_{(n_1,\cdots,n_k)}$ 
can be regarded as the crease set of the event horizon.

\section{An example of the black lenses}
\label{sec:kt}

Here we show an example of the topological model of the lens space black hole
in terms of a known exact solution of the Einstein equation.

\subsection{The Kastor-Traschen solution in five-dimensions}
The $d$-dimensional generalization of 
the Kastor-Traschen solution~\cite{Kastor:1992nn}
for Einstein-Maxwell equation with a positive cosmological term
is determined by the $(d-1)$-dimensional Riemannian Ricci flat metric $h$
on the base space
and the locations and masses of point sources on the base space.
Here we take as $h$ the Gibbons-Hawking multi-Taub-NUT metric~\cite{Gibbons:1979zt}
for $k$ linearly aligned nuts and 
locate the point sources on the nuts.
Ishihara {\em et.al.}~\cite{Ishihara:2006ig}
 have pointed out that this solution describes the coalescence of spherical black holes
into a lens space  black hole in the sense of the apparent horizon.
The purpose here is to clarify the topological structure of the event horizon
for the coalescing black holes.

Concretely, we consider the 5-dimensional Kastor-Traschen
solution with the metric in a local form
\begin{eqnarray}
\label{eq:5KT}
  g&=&-\dfrac{d\tau^2}{\left[-\sqrt{2\Lambda/3}\tau+U(\rho,z)\right]^2}
  +\left[-\sqrt{2\Lambda/3}\tau+U(\rho,z)\right] h\\
h&=&
\dfrac{(d\psi+w(\rho,z)d\phi)^2}{V(\rho,z)}
+V(\rho,z)(d\rho^2+d\rho^2+\rho^2 d\phi^2)\nonumber\\
U(\rho,z)&=&\sum_{i=1}^k \dfrac{m_i}{\sqrt{\rho^2+(z-z_i)^2}}\nonumber\\
V(\rho,z)&=&\sum_{i=1}^k \dfrac{1}{\sqrt{\rho^2+(z-z_i)^2}}\nonumber\\
w(\rho,z)&=&\sum_{i=1}^k \dfrac{z-z_i}{\sqrt{\rho^2+(z-z_i)^2}}\nonumber
\end{eqnarray}
where $\tau<0$ is the time coordinate, $\psi$ is the periodic coordinate,
of which period specified later,
 and $(\rho,z,\phi)$ is
the cylindrical coordinates.
The Maxwell field is determined by the $U(1)$ gauge field
\begin{eqnarray*}
  A=-{\sqrt{3}\over 2}\dfrac{d\tau}{\left[-\sqrt{2\Lambda/3}\tau+U(\rho,z)\right]}
\end{eqnarray*}
where $m_i>0$ $(i=1,\cdots,k)$ is the mass parameter of the
black hole on the $i$-th nut
and $z_i={\rm const}.$ 
$z_1<z_2<\cdots<z_k$ represent the location of the nuts aligned 
on the symmetric axis $\rho=0$ of the cylindrical coordinates.
This solution satisfies the
Einstein equation
\begin{eqnarray*}
  R_{\mu\nu}-{R\over 2}g_{\mu\nu}-\Lambda g_{\mu\nu}
=2F_{\mu\lambda}F_{\nu}{}^\lambda-{1\over 2}F_{\lambda\rho}F^{\lambda\rho}g_{\mu\nu}
\end{eqnarray*}

\subsection{The topology of the Gibbons-Hawking space}
The Gibbons-Hawking space appeared above, which has the metric $h$ by definition,
has Dirac-Misner string singularities on $\rho=0$.
To remove these singularities,
the Gibbons-Hawking space 
with $k$ nuts removed
should be regarded as a circle bundle 
over the  Euclidean 3-space $E^3$ minus $k$ points.
Let $M^\circ_{GH}$ denote the Gibbons-Hawking space minus $k$ nuts.
We define 
the projection map $\pi: M^\circ_{GH}\to E^\circ$
by
\begin{eqnarray*}
\pi: (\rho,z,\phi,\psi)\mapsto (\rho,z,\phi)
\end{eqnarray*}
where $E^\circ=E^3-\bigcup_i \{p_i\}$
is the
Euclidean 3-space
minus $k$ points $p_i:(\rho,z)=(0,z_i)$, $(i=1,\cdots,k)$.


We take the
$k+1$ coordinate neighborhoods of $E^\circ$ 
defined by
\begin{eqnarray*}
  U_0&=&E^3-\{\rho=0,z_1\le z\}\\
  U_i&=&E^3-\{\rho=0,z\le z_i\}-\{\rho=0,z_{i+1}\le z\}\quad (1\le i\le k-1)\\
 U_k&=&E^3-\{\rho=0,z\le z_k\}
\end{eqnarray*}
The local parametrization of $\pi^{-1}(U_a)$ $(a=0,\cdots,k)$ of 
this circle bundle is given by 
the new fibre coordinate $\psi_a$ defined by
\begin{eqnarray*}
  \psi_0&=&\psi-k\phi\\
\psi_1&=&\psi-(k-2)\phi\\
&\vdots\\
\psi_a&=&\psi-(k-2a)\phi\\
&\vdots\\
\psi_k&=&\psi+k\phi
\end{eqnarray*}
Then the Dirac-Misner string singularities on the axis
are all removed.
The transition of fibre coordinates between $\pi^{-1}(U_a)$ and $\pi^{-1}(U_{b})$
is given by
\begin{eqnarray*}
  \psi_{b}=\psi_a+2(b-a)\phi
\end{eqnarray*}
This shows that the transition functions take values in $U(1)$ if
the fibre coordinate $\psi_a$ is  identified  modulo $4\pi$ for all $a=0,1,\cdots,k$.

Identifying every $\psi_a$ modulo $4\pi$ in this way, the metric $h$ behaves 
near the $i$-th nut like the flat metric
$
h\sim dR^2+R^2 g(S^3),
$
where $R=[\rho^2+(z-z_i)^2]^{1/4}$ and $g(S^3)$ denotes the standard Riemannian metric 
on the unit 3-sphere.
Hence 
the Gibbons-Hawking space $M^\circ_{GH}$ is regularly extended to 
the full Gibbons-Hawking space $M_{GH}=M^\circ_{GH}\cup_i\{p_i\}$ including nuts.

\begin{figure}[htbp]
\centerline{\includegraphics[width=.5\linewidth]{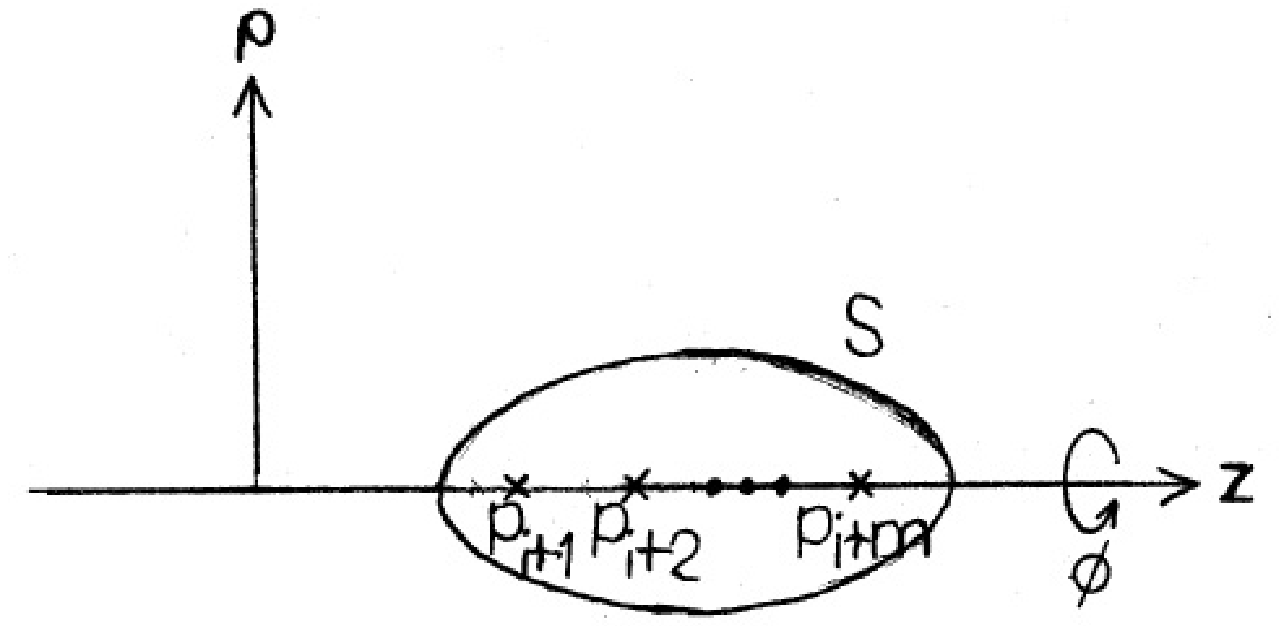}}
\caption{}
\label{fig:13_S2_in_MGH}
\end{figure}

Let us first note the structure of several submanifolds of the Gibbons-Hawking space $M_{GH}$.
The orientation of $M_{GH}$ is fixed such that
the local coordinates $(\rho,\phi,z,\psi_a)$ are positively ordered.
Let $S\subset E^\circ$ be a smooth 2-sphere enclosing $m$ points $p_{i+1},\cdots,p_{i+m}$
in $E^3$ as in 
Fig.~\ref{fig:13_S2_in_MGH}.
The restriction of $M^\circ_{GH}$ over $S$ is a circle bundle over the 2-sphere
and hence it is realized as the boundary of a 2-disk bundle over the 2-sphere described
in Sec.~\ref{sec:lp1}.
The 2-sphere $S$ is covered by two coordinate patches $U_i$ and $U_{i+m}$.
The transition between the fibre coordinates is 
\begin{eqnarray*}
  {\psi_i\over 2}={\psi_{i+m}\over 2}-m\phi
\end{eqnarray*}
Comparing this with the Eq.~(\ref{eq:attaching_map(p,1)}),
the total space of the circle bundle over $S$ turns out to be diffeomorphic
with $\partial M_{m}\simeq L(m,1)$.
In other words, the neighboring $m$ nuts are enclosed by the lens space $L(m,1)$
embedded in $M_{GH}$.
In particular, this becomes $S^3\simeq L(1,1)$, when $m=1$.
This reflects 
our requirement that every nut is a regular manifold point.

Next let us consider the subset $S_i:\{\rho=0,z_i\le z\le z_{i+1}\}$ $(i=1,\cdots,k-1)$
of $M_{GH}$, each of which is often called the {\em bolt}.
It is clear that $S_i$ is a topological 2-sphere, for
$S_i-\{p_i,p_i+1\}$ is an open cylinder. Furthermore, it can be easily
shown that the 2-sphere $S_i$ is smoothly
embedded into $M_{GH}$ for all $i$.
Let $N(S_i)$ denotes the tubular neighborhood of $S_i$ in $M_{GH}$.
In general, the tubular neighborhood of a 2-sphere embedded in a 4-manifold
is a 2-disk bundle over the 2-sphere. 
Since $N(S_i)$ contains neighboring two nuts, its boundary must be $L(2,1)$
which is diffeomorphic with the real projective 3-space $P^3(\R)$.
Hence $N(S_i)$ must be either $M_2$ or $M_{-2}$. 
Note that $\partial M_{-2}\simeq L(-2,1)\simeq L(2,1)$ under the orientation
preserving diffeomorphism. 

Whether $N(S_i)$ is $M_2$ or $M_{-2}$ is determined for example 
by considering the self-intersection number of $S_i$ as follows.
In general, two closed 2-submanifolds $S$ and $S'$
in a 4-manifold will intersect at a finite number of points.
Let $e^1\wedge e^2$ and $e^{1'}\wedge e^{2'}$ be the area form of $S$ 
and $S'$ at the intersection point, respectively.
The index of the intersection point is defined to be $1$ or $-1$
according to $e^1\wedge e^2\wedge e^{1'}\wedge e^{2'}$ is positively or
negatively oriented.
Then, the {\em intersection number} between oriented closed 2-submanifolds
$S$ and $S'$ is defined to be the sum of
the indexes over all intersection points.
The {\em self-intersection number} of a closed 2-submanifold $S$ is
defined to be the intersection number between $S$ and its continuous distortion
$S'$, where $S$ and $S'$ are arranged such that they intersect at a finite number of points
as always possible.
Note that the self-intersection number does not depend on the choice of the
orientation of $S$.
It can be easily seen that the self-intersection number of $S_i$ in $N(S_i)$ is
$-2$. While in general the self-intersection number of the zero section of $M_p$ is 
given by $p$.
Hence $N(S_i)$ must be $M_{-2}$.

The 2-spheres $S_i$ and $S_{i+1}$ $(i=1,\cdots,k-2)$ intersect at the $(i+1)$-th nut $p_{i+1}$.
The intersection number between them turns out to be $1$. (or $-1$ according to the choice
of the orientations of $S_i$ and $S_{i+1}$.)
This implies that the union $N(S_i)\cup N(S_{i+1})$  is  the plumbing 
of two $M_{-2}$, and $S_i\cup S_{i+1}$ gives its frame $\cup^2 B$.

Now it is clear that the union $N(S_i)\cup N(S_{i+1})\cup \cdots \cup N(S_{i+m-2})$
is the plumbing of $m$ 2-disk bundles $M_{-2}$.
Hence its boundary is $L(m,-m+1)\simeq L(m,1)$, which is
computed as
\begin{eqnarray*}
(-2,-2,\cdots,-2,-2,-2)\to
\left(-2,-2,\cdots,-2,-{3\over 2}\right)\\
\to
\left(-2,-2,\cdots,-{4\over 3}\right)\to
\left(-2,-{m-1\over m-2}\right)\to
\left(-{m\over m-1}\right)
\end{eqnarray*}
in terms of the rational blow down.
This is consistent with the fact that $N(S_i)\cup N(S_{i+1})\cup \cdots \cup N(S_{i+m-2})$
contains neighboring $m$ nuts.

In particular, the Gibbons-Hawking space $M_{GH}$ is diffeomorphic with
the interior of $N(S_1)\cup \cdots \cup N(S_{k-1})$, so that it can be written as
\begin{eqnarray*}
M_{GH}\simeq   {\rm int} M_{(-2,\cdots,-2)}\quad (\mbox{plumbing of $k-1$ copies
of $M_{-2}$})
\end{eqnarray*}
for $k\ge 2$.
This fact is pointed out in Ref.~\citen{Anderson:1989}.

\subsection{Formation of the lens space black holes}
The Kastor-Traschen space-time describes the merging of several black holes
in the de Sitter-like background.
This is seen by investigating the apparent horizons on  $\tau={\rm const.}$
hypersurfaces. That the 5-dimensional
Kastor-Traschen solution with the Gibbons-Hawking base space
actually represents the merging of black holes
is confirmed in a numerical way by Yoo {\em et.al}~\cite{Yoo:2007mq}
in the case of $k=2$,
where they show that it describes the merging of two $S^3$ black holes into 
a $L(2,1)$ black hole.

It is plausible that the Kastor-Traschen space-time
with the metric (\ref{eq:5KT}) for general $k$
describes the merging black holes 
in the sense of the event horizons.
An evidence is that this space-time approaches to 
the extreme Reissner-Nordstr\o m-de Sitter space-time with a $S^3/\Z_k\simeq L(k,1)$ horizon
at late time ($\sqrt{\Lambda}|\tau|\ll 1$), which describes a single static 
charged black hole with a regular event horizon. 
To determine the precise location of the event horizon, 
we will need numerical computations.
However, by symmetry consideration, 
we can extract essential features of the event horizon.
The following argument should be regarded as the most plausible scenario for 
the topology change of black holes.

\begin{figure}[htbp]
  \centerline{\includegraphics[width=.5\linewidth]{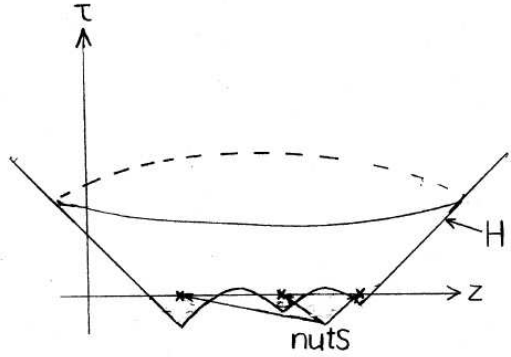}}
\caption{The Kastor-Traschen space-time represents the coalescence of several black holes.}
\label{fig:14_Kastor-Traschen}
\end{figure}

Since the space-time described by the
metric (\ref{eq:5KT}) admits the $U(1)\times U(1)$ isometry, which consists of
the translations in the $\phi$ and $\psi$--directions,
the event horizon is also invariant under this isometry.
Then the location of the event horizon at the time $\tau$ will be
represented by curves in the $(\rho,z)$--plane. 
We are now interested in the event horizon formed at finite past.
So, let us regard these black holes as formed by gravitational collapses
occurred around the nuts 
(See Fig.~\ref{fig:14_Kastor-Traschen}).
We would be able to neglect the effect of the gravitational waves on
the topology of the event horizon.

\begin{figure}[htbp]
\centerline{\includegraphics[width=.5\linewidth]{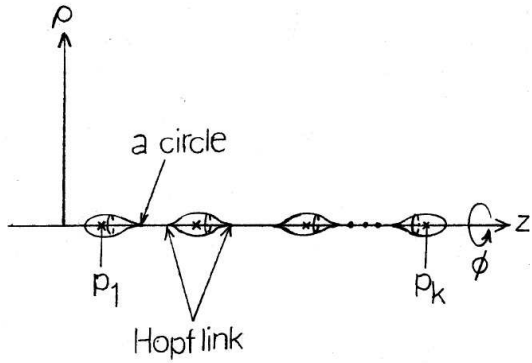}}
\caption{The snap shot of the event horizon. Each horizon around $p_1$ and $p_k$
has a cusp of a circle and every other horizon  has cusps forming a Hopf link.}
\label{fig:14.1_cusp_structure}
\end{figure}

At early time ($\sqrt{\Lambda}|\tau|\gg 1$), the space-time contains $k$ apparent horizons
with $S^3$ horizons,
each surrounding a point source at a nut.
Then, we expect that the event horizons are also formed slightly outside the
apparent horizons.
In other words, the event horizon at early time contains $k$ connected components
each homeomorphic with the 3-sphere.
Since these black holes will coalesce at late times, each black hole will have
cusps on the horizon. The $1$-st and $k$-th horizons
will have a single cusp
and the other horizons will have a pair of cusps.
All the cusps will be aligned on the $\rho=0$ axis, and they are all topologically
a circle. 
This expectation comes from the following consideration.
These cusps are the snap shot of the crease set at the time $\tau$.
By symmetry, the crease set can be regarded as being aligned on the $\rho=0$ axis.
Concretely, the crease set projected onto a $\tau={\rm const.}$ slice
will be represented by
the interval $\{\rho=0,z_1\le z\le z_k\}$, and hence it will be
the frame $\cup^{k-1}B=S_1\cup\cdots\cup S_{k-1}$ of a plumbing manifold.
The crease set looks like a 2-disk around $p_1$ and $p_k$,
and it looks like transversally intersecting 2-disks around $p_i$ for $2\le i\le k-1$.
Therefore, the snap shot of the crease set will consist of 
a single circle in both $S^3$ horizons around $p_1$ and $p_k$, 
and a Hopf link in every $S^3$ horizon around other nut 
(See Fig.~\ref{fig:14.1_cusp_structure}).

\begin{figure}[htbp]
\centerline{\includegraphics[width=.5\linewidth]{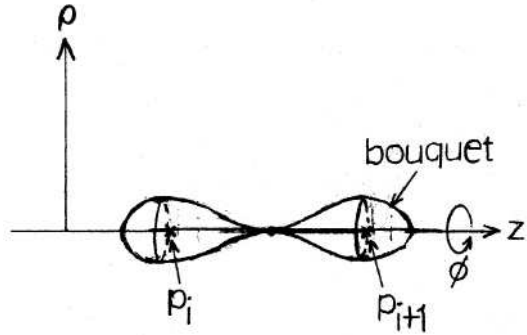}}
\caption{The snap shot of the event horizon when two black holes merge.
Projected on $E^3$, the horizon is a bouquet of two 2-spheres.}
\label{fig:15_bouquet}
\end{figure}

\begin{figure}[htbp]
\centerline{\includegraphics[width=.5\linewidth,height=.3\textheight]{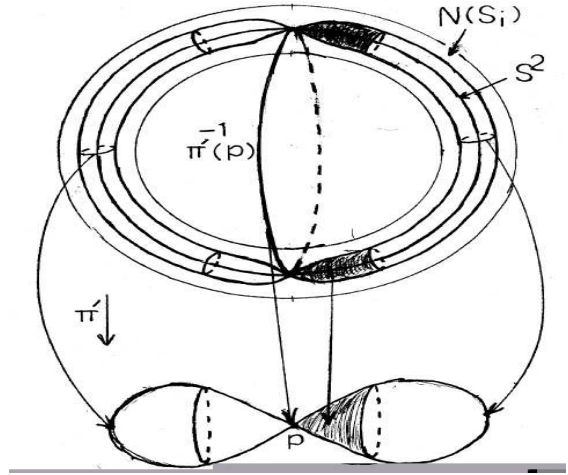}}
\caption{The embedding of the circle bundle over the bouquet into the 2-disk bundle $N(S_i)$
over $S^2$ is schematically depicted. The projection map of the circle bundle is denoted
by $\pi'$. The shaded region in the circle bundle
is homeomorphic with the solid torus, and it is projected
onto the shaded region of the bouquet. The fibre over $p$ corresponds to the equator
of the $S^2$ base space of $N(S_i)$.}
\label{fig:16_S1_bundle}
\end{figure}

\begin{figure}[htbp]
\centerline{\includegraphics[width=.5\linewidth]{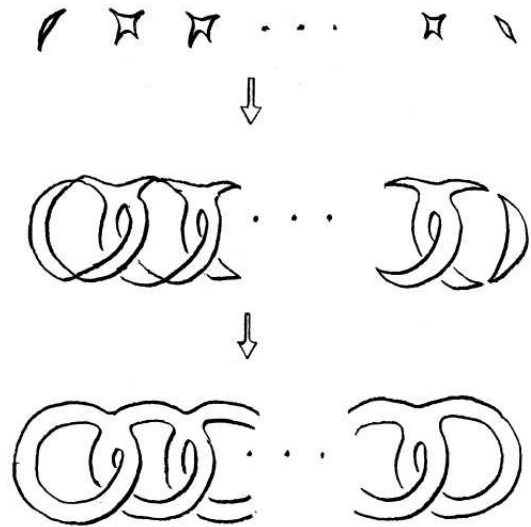}}
\caption{The time evolution of the event horizon is schematically depicted,
where two of the four spacial dimensions are omitted.
The $k$ spherical horizons around nuts merge into a single $L(k,1)$ black lens.}
\label{fig:17_bl_formation}
\end{figure}

These black holes will coalesce at intermediate times.
Each coalescence will occur on the axis $\{\rho=0\}$.
If $i$-th and $i+1$-th horizons coalesce first, the snap shot 
at the coalescence time
of the horizon
will be a pair of 2-spheres touched at a point,
which we call a bouquet of two 2-spheres,
when projected onto $E^3$
(See Fig.~\ref{fig:15_bouquet}).
Hence this portion of the horizon will be the circle bundle over the bouquet.
How this is embedded in the 2-disk bundle $N(S_i)$ is depicted in 
Fig.~\ref{fig:16_S1_bundle}.
This bouquet will immediately transform into a 2-sphere enclosing two nuts $p_i$ and $p_{i+1}$
in $E^3$.
The event horizon at this time is the circle bundle over the 2-sphere, which is topologically
the lens space $L(2,1)$.
Repeating such elementary processes of the merging, there will be 
black holes with several connected components at a later time, each will be a topologically
lens space $L(m,1)$ for some $m\ge 1$, if it encloses $m$ nuts.
Finally, these black holes will merge into a lens space black hole with the $L(k,1)$
horizon
(See Fig.~\ref{fig:17_bl_formation}).

In this way, the 5-dimensional Kastor-Traschen solution with the Gibbons-Hawking
base space with $k$ nuts will describe the formation of the $L(k,1)$ black hole
via the merging of several lens space black holes.

\subsection{Black rings off the nuts}
The above description of the topology change of the black holes is based on the
time slicing with respect to the time function $\tau$.
If we choose another time slicing, the evolution of black holes may look differently.
This is the consequence of the fact that the crease set is an achronal set.

\begin{figure}[htbp]
\centerline{\includegraphics[width=.5\linewidth]{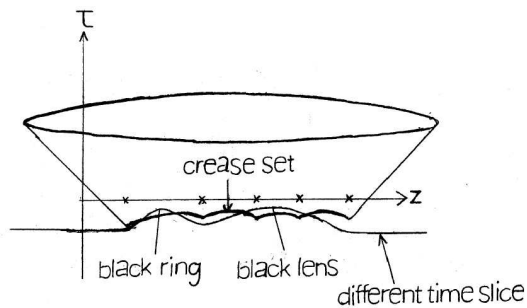}}
\caption{In a different time slicing, the space-time might describe a black ring formation.}
\label{fig:18_ring_lens}
\end{figure}

In the following, we consider a different time slicing
but still respecting the $U(1)\times U(1)$ isometry.
There is a time slicing, in which a black hole off the nuts first appear.
The black hole region at this moment will be a 3-disk 
when projected onto $E^3$ with its center on the $\rho=0$ axis.
The boundary 2-sphere corresponds to the time slice of the event horizon projected onto $E^3$,
and it will have two cusps on the axis.
Hence the event horizon at this moment is homeomorphic with $S^2\times S^1$,
namely a black ring (See Fig.~\ref{fig:18_ring_lens}).
This black ring will soon turn into an $S^3$ black hole, when it incorporates
a neighboring nut.

Even based on the time slicing with respect to $\tau$, we can consider 
the formation of black rings.
It is realized by considering a massive source off the nut.
In other words, we just add terms
\begin{eqnarray*}
\sum_{j=1}^l \dfrac{\mu_j}{\sqrt{\rho^2+(z-\zeta_j)^2}}\nonumber\\  
\end{eqnarray*}
to the metric function $U(\rho,z)$,
where $\zeta_j\ne z_i$ for all pairs of $j$ and $i$.
These terms represent $l$ massive circle sources.
It turns out that these sources correspond to space-time singularities. 
However, we can avoid this problem by considering 
that these circle sources appear only after a finite past
due to the gravitational collapses
 and these singularities are hidden behind the
event horizon.

Then, we can consider the merging of several black rings and several black lenses.
The elementary processes for the merging of black holes in this space-time can be
written in a loose expression as
\begin{eqnarray*}
  L(m,1)+L(m',1)\longrightarrow L(m+m',1)\quad (m,m'\ge 0)
\end{eqnarray*}
which represents a merging of the $L(m,1)$ horizon and the $L(m',1)$ horizon
into the $L(m+m',1)$ horizon.

\section{Conclusion}
\label{sec:conc}
We have considered the topological structure of the event horizon which evolve
into a black hole with a horizon diffeomorphic with the lens space.
We show such a black hole can emerge from a crease set with relatively simple structure,
which is constructed by connecting several 2-spheres.
Then, the event horizon is modeled by the plumbing manifold of several
2-disk bundles over the 2-sphere.
Hence a simple picture is obtained for the formation of the black lens, 
where the evolution of the black hole can 
be viewed as several 2-handle attachments.

It is also shown that the event horizon with such topology is realized in
the Kastor-Traschen solution of the Einstein-Maxwell theory. 
Although this solution only describes the black lens of the special type $L(k,1)$,
it might provide useful examples
of merging black rings and  black lenses.

There has never been known exact solutions of the Einstein equation
for black lenses, which is asymptotically flat nor has a general $L(p,q)$ horizon.
If we have a Ricci flat metric on a plumbing 4-manifold,
a simple example of $L(p,q)$ black lens 
similar to that in this paper will be obtained
by taking the Ricci flat space as the base space of the Kastor-Traschen space-time.

The spacial section of the domain of outer communication
in an asymptotically flat black lens space-time will be given by
a cobordism between the lens space and the 3-sphere. It should also be simply connected
as required by the topological censorship theorem. 
An example is given by the plumbing 4-manifold $M_{(n_1,\cdots,n_k)}$ minus $D^4$.
This 4-manifold is simply connected, for the deformation retract
$\cup^k B$ of $M_{(n_1,\cdots,n_k)}$ is simply connected, and it is a cobordism between $L(p,q)$ and $S^3$.



\end{document}